\documentclass{icrc2009}

\usepackage{graphicx}   % for including figures
\usepackage{caption}    % for captions
\usepackage[font=footnotesize]{subfig} % subfig.sty for a double column floating figure using two subfigures
\usepackage{fixltx2e}
\usepackage{url}
\usepackage{amssymb}
\usepackage{lineno}

\newcommand{\shorttitle}[1]%
{\markboth{Proceedings of the 31\MakeLowercase{$^{st}$} ICRC, {\L}\'{o}d\'{z} 2009}{#1} }
\newcommand{\etal}{\MakeLowercase{\textit{et al. }}} % "et al."

\begin{document}
\title{BATATA: A device to characterize the punch-through observed in underground muon detectors 
and to operate as a prototype for AMIGA}

\author{\IEEEauthorblockN{Medina-Tanco Gustavo\IEEEauthorrefmark{1},
			  for the Auger Collaboration\IEEEauthorrefmark{2}}
                            \\
\IEEEauthorblockA{\IEEEauthorrefmark{1}Instituto de Ciencias Nucleares (ICN), Univ. Nacional Aut\'onoma de M\'exico, M\'exico D.F.}
\IEEEauthorblockA{\IEEEauthorrefmark{2} Observatorio Pierre Auger, Av. San Mart\'{\i}n Norte 304 (5613) Malarg\"ue, Prov. Mendoza, Argentina.}}

% please write the preseter's name and short title (3-4 words maximum)
%    which will appear at the header of the even pages.
\shorttitle{Medina-Tanco G. \etal BATATA muon detector}
\maketitle

\begin{abstract}
BATATA is a hodoscope comprising three X-Y planes of plastic scintillation detectors. 
This system of buried counters is complemented by an array of 3 water-Cherenkov 
detectors, located at the vertices of an equilateral triangle with 200 m sides. 
This small surface array is triggered by extensive air showers. The BATATA detector 
will be installed at the centre of the AMIGA array, where it will be used to quantify 
the electromagnetic contamination of the muon signal as a function of depth, and so 
to validate, in situ, the numerical estimates made of the optimal depth for the AMIGA 
muon detectors.  BATATA will also serves as a prototype to aid the design of these 
detectors.  \end{abstract}

\begin{IEEEkeywords}
muon-hodoscope, punch-through-characterization, AMIGA
\end{IEEEkeywords}

%Insert linenumbers from here -- comments before submission 
%\linenumbers
 
\section{Introduction} \label{sec:intro}

High energy cosmic rays are indirectly characterized by measuring the extensive air shower cascades that they trigger in the Earth atmosphere. At ground level, and at distances greater than a few tens of meters from the axis, the shower is dominated by just two components: electromagnetic (electrons, positrons and photons) and muonic. The relative weight of these two components has invaluable information about the nature of the primary cosmic ray and of the high energy hadronic processes taking place at high altitude during the first interactions. 
Water Cherenkov detectors used by the Pierre Auger Observatory measure the combined energy deposition of charged particles inside its volume and are not well suited for discriminating between electromagnetic and muonic components. Therefore, as part of the Auger surface low energy extension AMIGA \cite{AMIGA}, buried muon scintillators will be added to the regular surface stations of an infill region of the array. Provided enough shielding is ensured, these counters will allow the measurement of the muon signal and, in combination with their Cherenov tank companions, the estimate of the electromagnetic component. The final objective of this assemblage is to achieve high quality cosmic ray composition measurements along the ankle region \cite{EnhancementsMerida2007}. 
Several prototyping activities are being carried out for the muon counters \cite{AMIGAPtotoypes} and an additional detector is being constructed, BATATA. The main objective of the latter is to quantify 
the electromagnetic contamination of the muon signal as a function of depth, and so 
to validate in situ the numerical estimates made of the optimal depth for the AMIGA 
muon detectors.  BATATA will be installed at the center of the future AMIGA array and will also 
 serve as a prototype to aid the design of its detectors. The chosen site, from the point of view 
 of soil properties and, therefore, of punch-through characteristics, is statistically equivalent to 
 any other location of the area covered by the infill \cite{LinkToGeoGAPAtICN}. 
\\
 
\section{The detector} \label{sec:detector}

\begin{figure}[!t]
\centering
\includegraphics [width=0.45\textwidth]{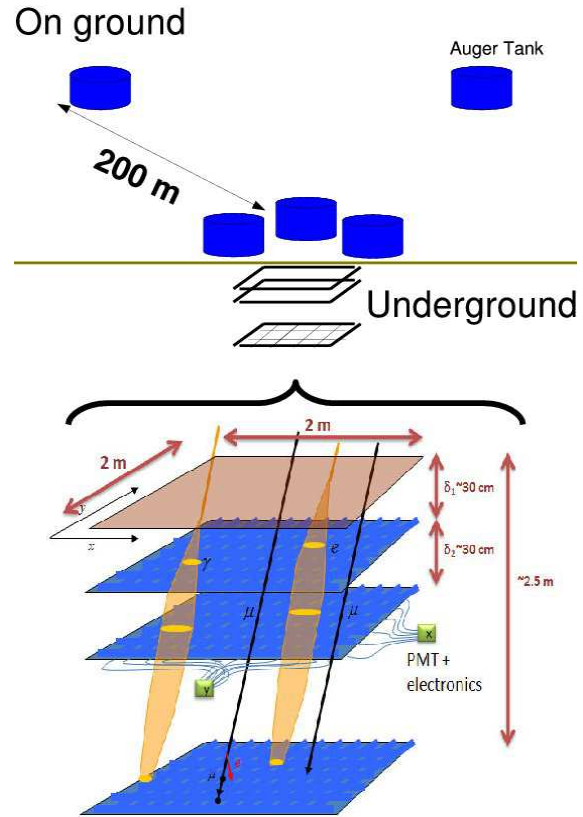}
%\caption{Buried section of the BATATA detector and working principle.}
\caption{Surface (top) and buried (bottom) sections of the BATATA detector and working principle.}
\label{BATATASchematics}
\end{figure}

A schematic view of BATATA is shown in Figure \ref{BATATASchematics}. The detector is 
composed by a set of three parallel dual-layer scintillator planes, buried at fix depths ranging 
from $0.50$ m to $2.5$ m.

\begin{figure}[h]
\centering
\includegraphics [width=0.48\textwidth]{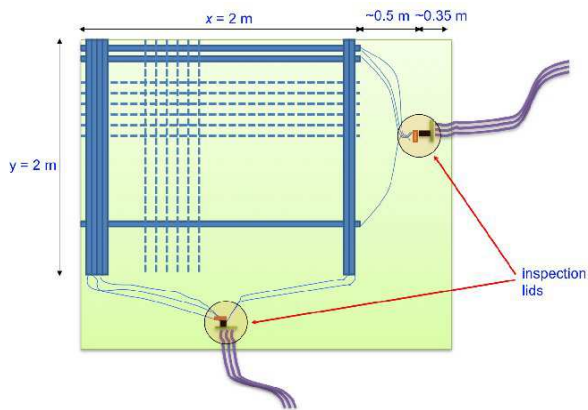}
\caption{
Schematic view of the arrange of scintillators, optical fibers and front-end electronics inside the casing for any one of the three BATATA planes.}\label{CasingSchematics}
\end{figure}

Each layer in a plane is 4 m$^{2}$ and is composed by $49$ rectangular strips 
of $4$ cm x $2$ m, oriented at a right angle with respect to its companion layer, which gives 
an xy-coincidenceÊpixel of 4x4 cm$^{2}$. The scintillators are MINOS-type extruded polystyrene 
strips \cite{MINOSScintillators}, with an embedded Bicron BC92 wavelength shifting fiber, 
of $1.5$ mm in diameter \cite{BicronBC92}. Building and quality assessment protocols  (e.g., cutting, polishing, gluing and painting) as well as the characterization procedures of the optical fibers 
and scintillator bars can be found in \cite{LinkToGapsOFChar}. 
Light is collected by Hamamatsu H7546B 64 pixels 
multi-anode PMTs \cite{PMTH7546B}.  Each $x$-$y$ plane is fitted inside an 
individual casing, where each orthogonal direction ends in its own front-end electronic board (FE),
as shown schematically in Figure \ref{CasingSchematics}.

The casings are made of fiber glass and were specially designed to be water- and light-tight 
and to withstand handling during shipping and burying. They have two caps which allow for 
easy access to the corresponding front-end boards and optical couplings between 
PMTs and optical fiber cookies. The caps can be dismounted in two steps if necessary, one for inspection and servicing, and a second for cabling replacement, ensuring in both cases that water-tightness 
can be recovered. Additionally, the casing and sealing are versatile enough to allow for the straight forward addition of new components and cabling not specifically foreseen in the original design. 

The front-end electronics works in counting mode and signals are transmitted to the surface DAQ 
stage using low-voltage differential signaling (LVDS). Any strip signal above threshold opens a 
GPS-tagged $2$ $\mu s$ data collection window. Data, including signal and background, are 
acquired by a system of FPGA Spartan boards and a TS7800 single board computer. 
The code controlling the data flux at the FPGA level was written in VHDL (VHSIC -Very High Speed Integrated Circuits- hardware description language). 

\begin{figure}[h]
\centering
\includegraphics [width=0.48\textwidth]{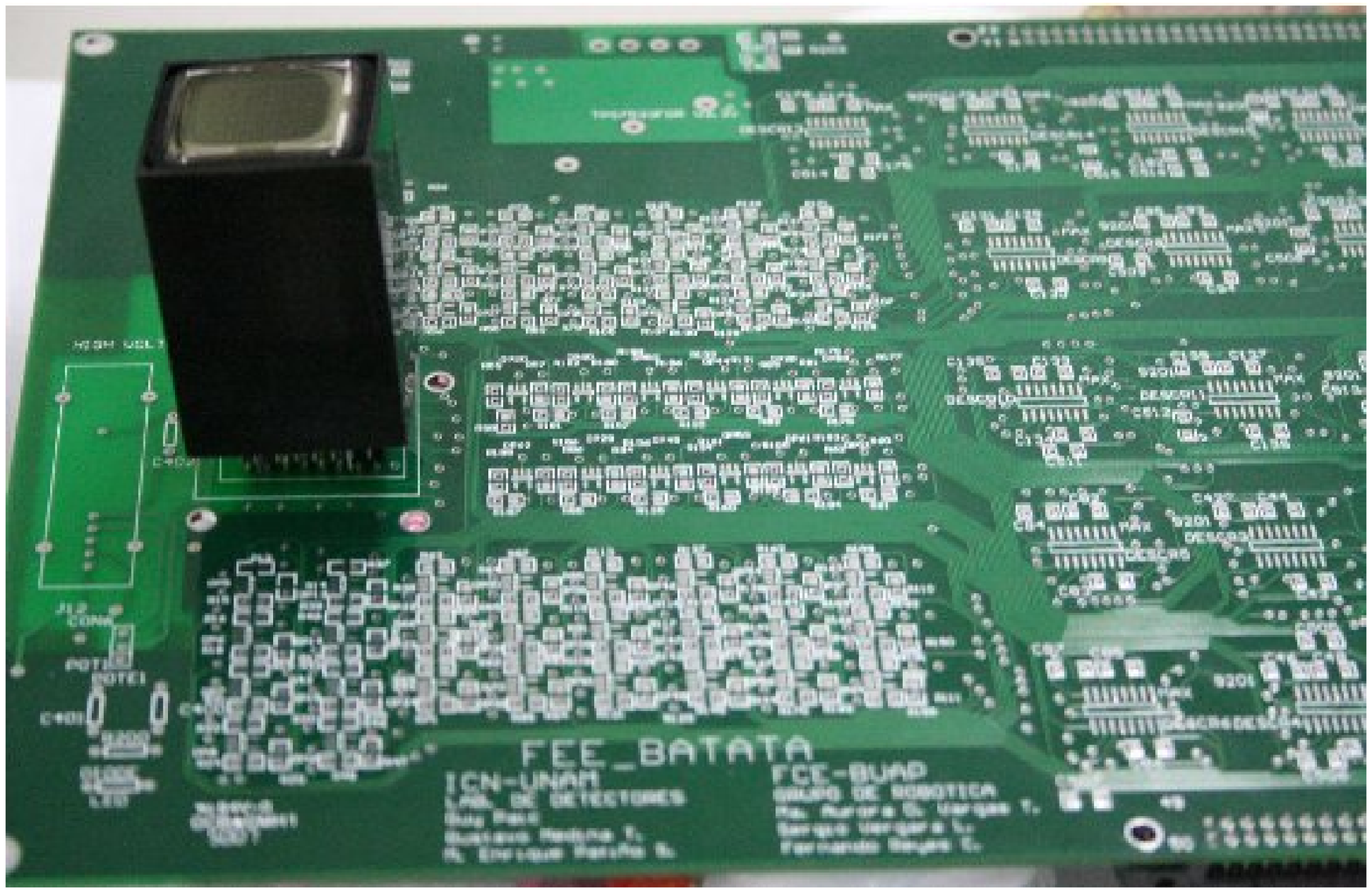}
\includegraphics [width=\columnwidth]{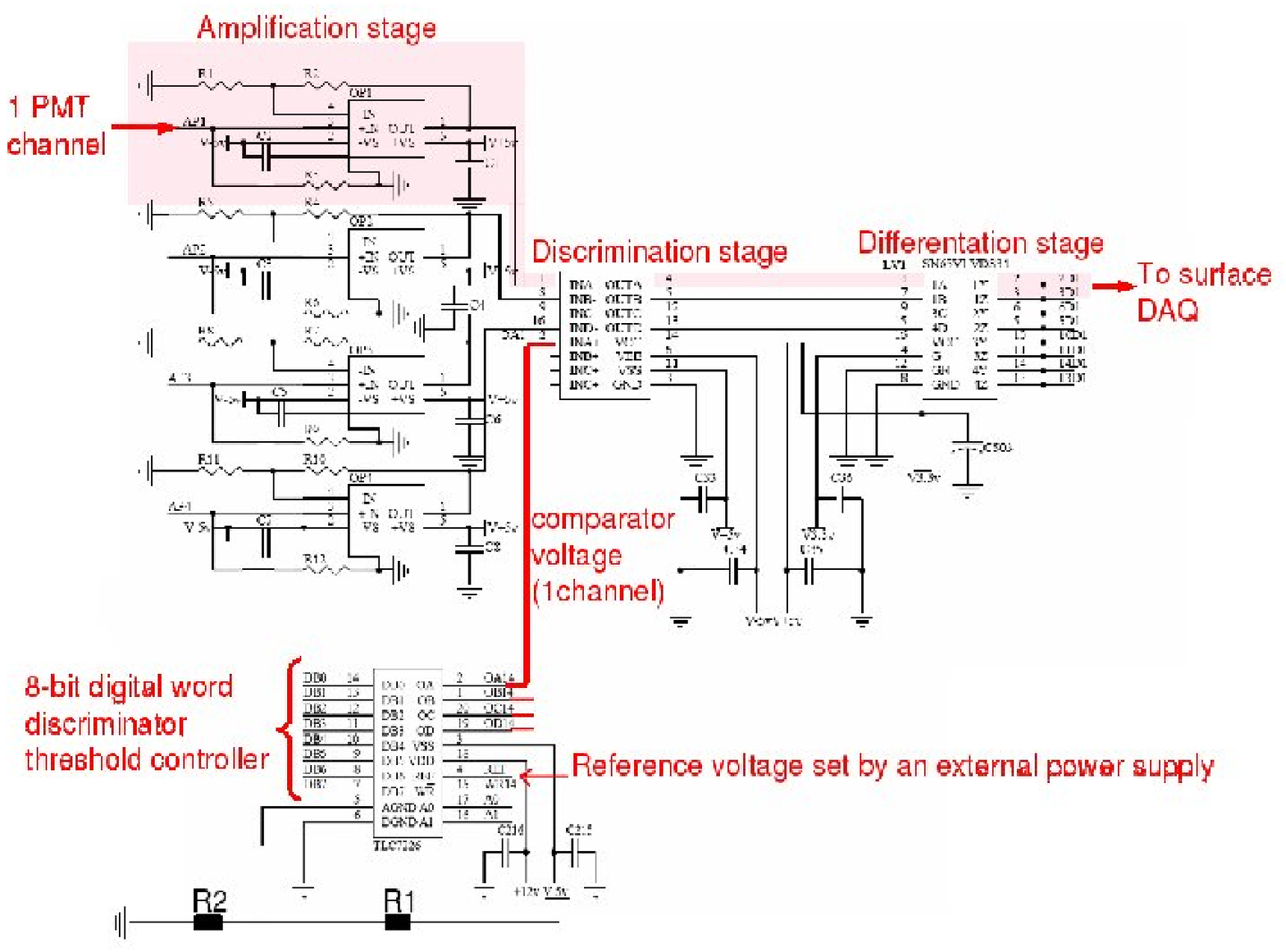}
\caption{Front-end board: (top) general view with multi-anode PMT, (bottom) single channel layout.}\label{FrontEnd_SingleChannel}
\end{figure}

The front end boards are $12.3'' \times 9.3''$ and comprise $64$ channels, of which only $49$ are actually used at present. The multi-anode PMT and its high-voltage supply are located on the board. Each channel includes: (i) an amplification stage, which uses the AD8009 operated at amplification factor $\sim 6$, (ii) a discrimination stage, which uses a MAX9201 high-speed, low power, quad comparator with fast propagation delay (7ns typ at 5mV overdrive) connected in bipolar mode, (iii) a digital-to-analog converter TLC7226C to independently setup the discrimination voltage of each channel and, (iv) a high-speed differential line driver SN55LVDS31 to transform the discriminator output into a differential signal which is carried out to the surface and to the DAQ system some $17$ m away (see, Figure \ref{FrontEnd_SingleChannel}).

\begin{figure}[h!]
\centering
\includegraphics [width=0.48\textwidth]{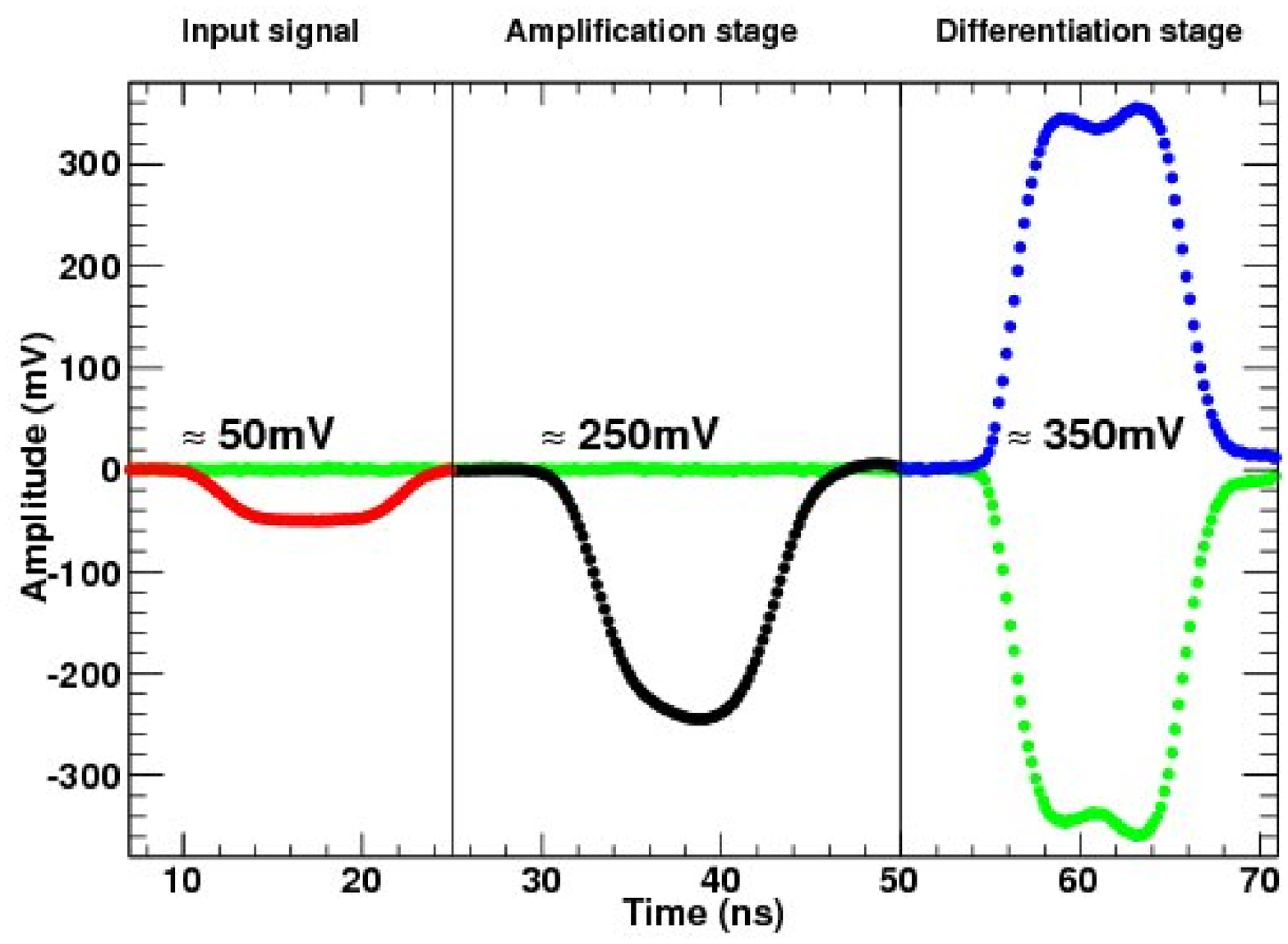}
\includegraphics [width=0.48\textwidth]{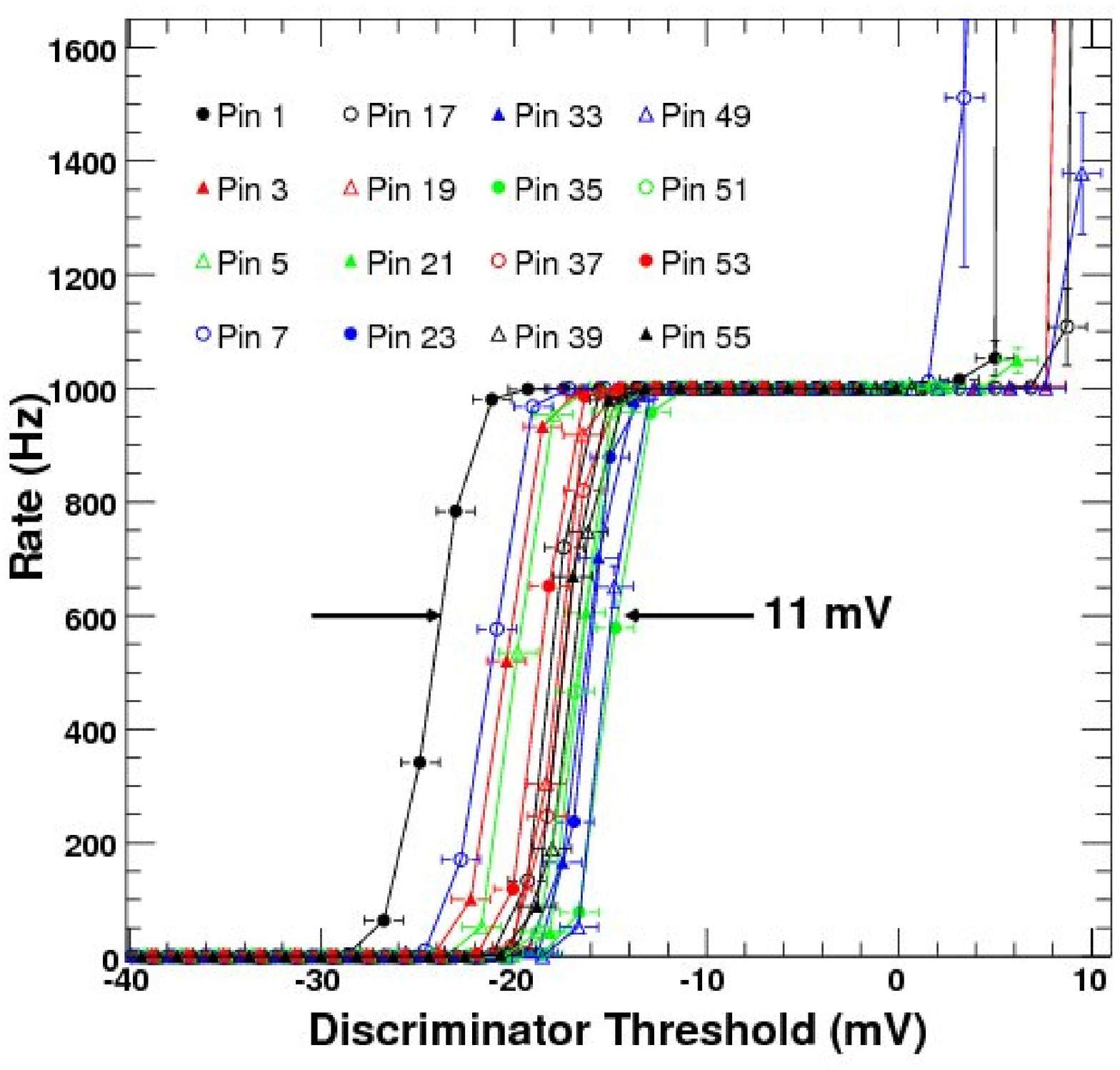}
\caption{Testing and characterization measurement on FE board with controlled square pulses of 50 mV height and $\approx$10 ns width injected at a frequency of 1 kHz.}\label{Square_pulses_stages_And_rates}
\end{figure}

%Mean pulse shape (mV vs. ns) and pulse hight (mV), rise time, T50-T10 (ns) and width, T90-T10 (ns) 
%Distribution functions for muons measured at 160 cm from the PMT window can be seen in Figure \ref{FrontEnd_pulses}.
%
%\begin{figure}[h!]
%\centering
%%\includegraphics [width=0.48\textwidth]{FrontEnd_pulses.eps}
%\caption{Muon pulses at 160 cm from the PMT window: (a) mean pulse shape (mV vs. ns), (b) pulse hight distribution function (mV), (c) rise time, T50-T10 (ns) and (d) width, T90-T10 (ns).}\label{FrontEnd_pulses}
%\end{figure}
Each stage of the FE board has been tested with controlled square pulses of 50 mV height and $\approx$10 ns width injected at a frequency of 1 kHz. 
Mean outputs of the amplification and differentiation stages as well as output rates of 16 different channels are shown in 
Figure \ref{Square_pulses_stages_And_rates}.

\begin{figure*}[!t]
%   \centerline{\subfloat[]{\includegraphics[width=2.5in]{E2ESim_MuonTracks.eps} \label{E2ESim_MuonTracks}}
%              \hfil
%              \subfloat[]{\includegraphics[width=2.5in]{E2ESim_Edeposit.eps} \label{E2ESim_Edeposit}}
%             }
%   \centerline{\subfloat[]{\includegraphics[width=2.5in]{E2ESim_PhotonsAtPMT1.eps} \label{E2ESim_PhotonsAtPMT1}}
%              \hfil
%              \subfloat[]{\includegraphics[width=2.5in]{E2ESim_PhotonsAtPMT2.eps} \label{E2ESim_PhotonsAtPMT2}}
%             }
   \centerline{\subfloat[]{\includegraphics[width=2.5in]{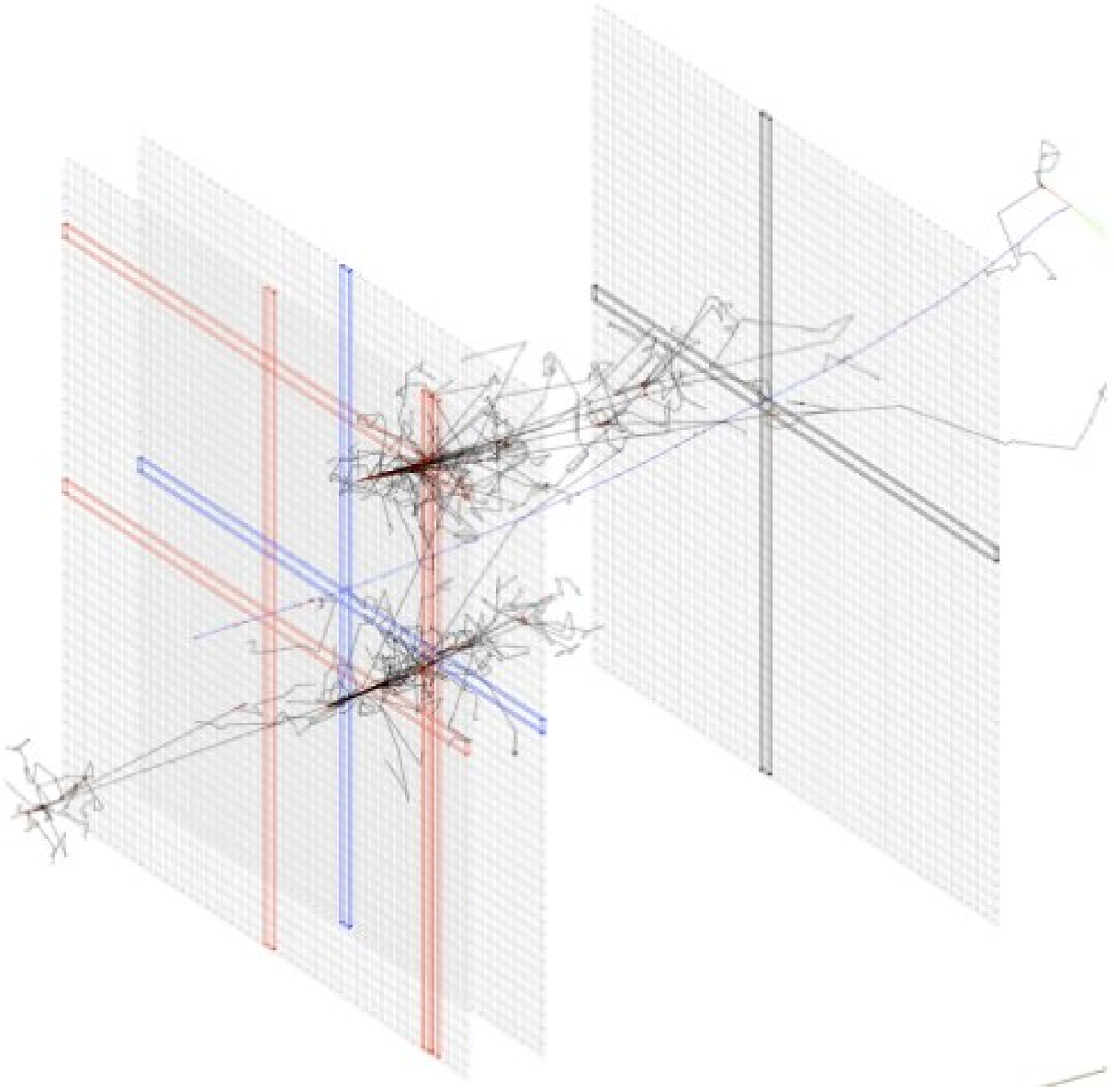} \label{E2ESim_MuonTracks}}
              \hfil
              \subfloat[]{\includegraphics[width=2.5in]{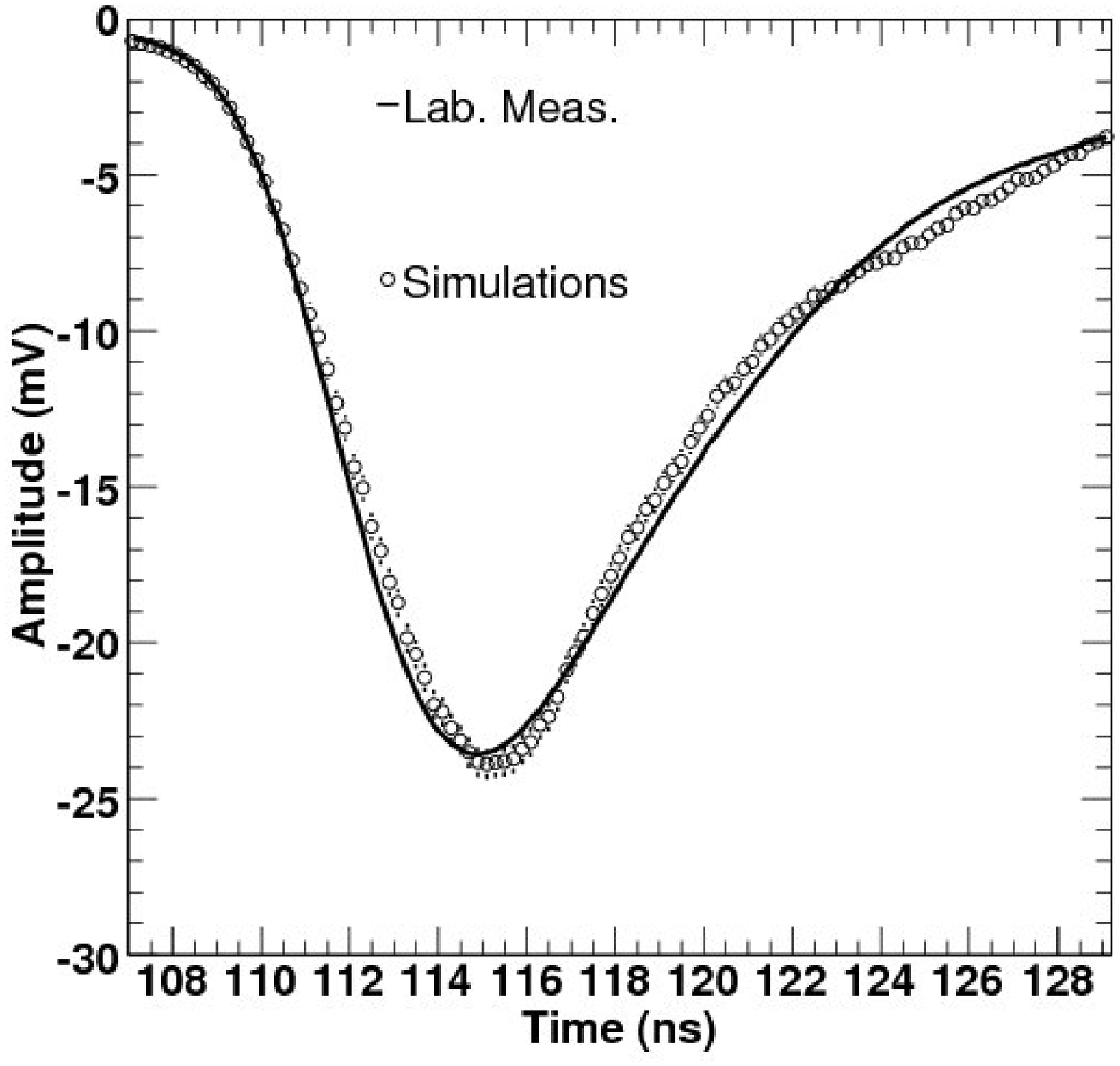} \label{E2ESim_Edeposit}}
             }
   \centerline{\subfloat[]{\includegraphics[width=2.5in]{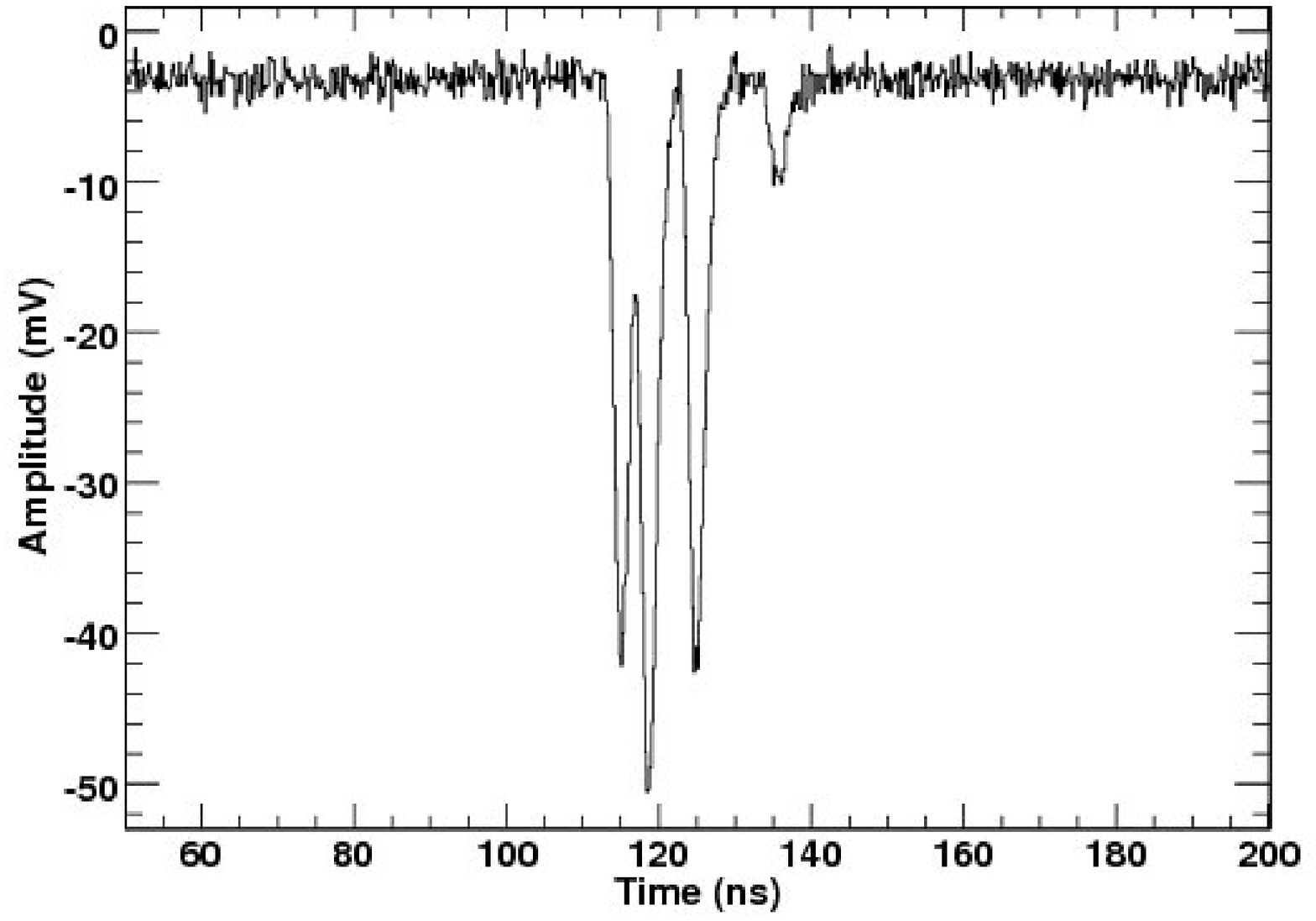} \label{E2ESim_PhotonsAtPMT1}}
              \hfil
              \subfloat[]{\includegraphics[width=2.5in]{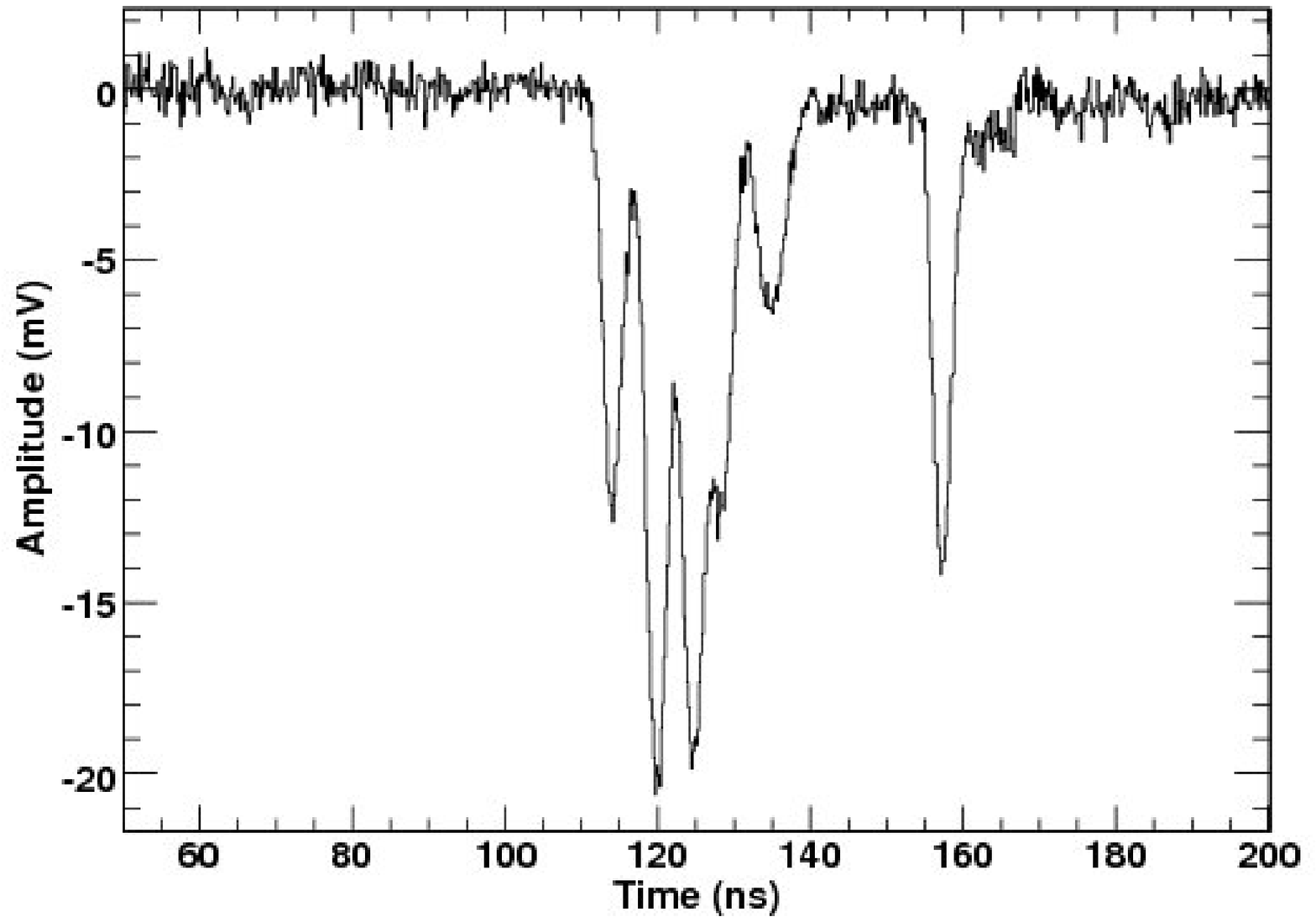} \label{E2ESim_PhotonsAtPMT2}}
             }
   \caption{End-to-end simulations. (a) Simulated muon tracks and low energy electromagnetic cascades. 
(b) Real and simulated mean pulse shape at 160 cm from the edge of a scintillator. 
(c) Simulated muon signal.
(d) Actual laboratory measured muon signal.}
   \label{E2ESim}
 \end{figure*}

The front end electronics resides in an enclosure which only exchanges heat with the surrounding 
ground by conduction. It has being experimentally determined that, under the most unfavorable 
conditions, the equilibrium temperature at the surface of the board never exceeds $40$ $^{o}$C,
which is well inside the working range of the electronic components.

A cosmic ray EAS event triggers at least tens of channels in a time scale o a $\mu$-sec. This rate is much higher than the $\sim$kHz background of the detector due to low energy cosmic rays and natural radioactivity. Therefore, EAS events can be easily detected with a simple trigger scheme implemented at the FPGA level. The FPGA trigger is implemented via software and can be easily changed if appropriate.
In order to ensure the stability of the trigger rate, background events are also recorded at fixed time intervals during each run and are used to re-calibrate the discriminator thresholds at the end of each run.
Additionally, in order to characterize the electromagnetic punch-through, EAS inside a limited energy range and with axis at a defined distance from the detector have to be selected. In order to attain this, BATATA also counts with a small triangular surface array of 3 regular SD Cherenkov stations at a separation of $200$ m. This array provides the required offline selection capability for quasi-vertical showers in the vicinity of $10$ PeV.   

The power consumption of BATATA is $\lesssim 200$ W. This power is supplied by an array of $20$ solar panels with their corresponding batteries, ensuring continuous operation during low insolation winter months.

\section{End-to-end simulations and data analysis}

Muons are penetrating particles and propagate through the ground following a well defined track. Therefore, it is expected that, most of the times, they will only trigger one pixel at each layer. Furthermore, muons will tend to trigger $x$-$y$ pixels in coincidence inside the same plane. Electrons, positrons and photons, on the other hand, are much less penetrating and generate rapidly evolving electromagnetic showers underground which have a lateral extension. Therefore, they will tend to leave 2-dimensional footprints in the detector, specially at the depths of maximum development of the cascades between $\sim 30$ and $\sim 80$ cm for particles that, on the ground, have energies in the range $\sim 0.5$--$10$ GeV. These differences between muon and electromagnetic signatures at the detector are used for discrimination.
This is schematically shown in Figure \ref{BATATASchematics}, and motivates the uneven depths chosen for the three scintillator planes of BATATA. 

The data analysis requires a thorough understanding of the statistical response of the detector to EAS particles impinging the ground above and in its vicinity. Therefore, a comprehensive set of numerical end-to-end simulations of the detector have been carried out. A combination of AIRES \cite{AIRES} and Geant4 \cite{Geant4} is used to simulate EAS development from the top of the atmosphere up to the ground, where particles are injected and followed using Geant4 while they propagate through the soil and into the scintillator bars. The subsequent photons produced by the dopants inside the polystyrene strips are followed while bouncing off the TiO$_{2}$ covering and into the wavelength shifting optical fiber. Finally, the green-shifted photons thus produced by the dopant of the fiber are transfered along up to the window of the PMT. The several parameters involved in the simulation of the scintillator are then fine-tunned to reproduce the structure of the time profiles and integrated charges of muon pulses actually measured under laboratory conditions. In parallel, the shower impinging the surface array is reconstructed using the Offline Auger reconstruction and data handling package \cite{Offline}. 

Figure \ref{E2ESim} shows some example of the output of the end-to-end simulations for the underground segment of BATATA. Figure \ref{E2ESim_MuonTracks} shows one simulated muon tracks and 2 low energy electromagnetic cascades crossing the 3 planes of the detector and the corresponding triggered strips, i.e., producing a voltage above threshold in their front-end discriminator. 
Figure \ref{E2ESim_Edeposit} shows the simulated and the real mean pulse shape for background muons.
%energy deposition at distances of $20$ , $100$ and $180$ cm from the edge of an individual scintillator strip versus the corresponding optical photons at the PMT window after propagation through the wavelength shifting optical fiber. 
The output of the overall simulation chain is an electronic PMT signal as shown in Figure \ref{E2ESim_PhotonsAtPMT1}.
As can be seen from Figure \ref{E2ESim_PhotonsAtPMT2}, where a measured muon signal is shown, 
%the simulated distribution function of the number of photons arriving at the PMT window from hits at the previous locations in the strip when the scintillator is under the unshielded flux of atmospheric muons. The actual measurement of these distributions in laboratory are shown in Figure \ref{E2ESim_PhotonsAtPMT2},
%showing that besides a normalization factor, 
the simulations are able to reproduce satisfactorily the output of the detector.

%\begin{figure}[t]
%\centering
%\includegraphics [width=0.48\textwidth]{E2ESim_Scheme.eps}
%\caption{End-to-end simulations of the detector include in an integrated way the EAS cascade in the atmosphere and the response of the surface array and its data reconstruction, as well as the propagation of muons and electromagnetic cascades through the ground and the consequent light production at the PMT pixels for each triggered channel.}\label{E2ESim_Scheme}
%\end{figure}

\section{Conclusions}
 
A hodoscope, BATATA, comprising three X-Y planes of plastic scintillation detectors, complemented 
by an array of 3 water-Cherenkov detectors at a separation of 200 m is under construction and will 
be installed at the centre of the AMIGA extension to the Auger baseline design. BATATA will be used 
to quantify the electromagnetic contamination of the muon signal as a function of depth. and will also 
serve as a prototype to aid the design of the AMIGA muon detectors.

BATATA is in its final phase of construction and its deployment will start on July 2009 with commissioning along the second semester of the year.
 
\section*{Acknowledgements}
This work is partially supported by the Mexican agencies CONACyT and UNAM's CIC and PAPIIT.

\end{document}